\documentclass[reprint,twocolumn,aps,superscriptaddress]{revtex4}
\usepackage{amssymb}
\usepackage{amsmath}
\usepackage{graphicx}
\usepackage[normalem]{ulem}
\usepackage[dvips]{color}
\usepackage{graphicx}
\usepackage{subfigure}

\renewcommand\sout{\bgroup \color{red} \ULdepth=-.5ex \ULset}

\begin{document}

\title{Pion production and absorption in heavy-ion collisions}
\author{Yuan Gao}
\email{gaoyuan@impcas.ac.cn}

\affiliation{School of Information Engineering, Hangzhou Dianzi
University, Hangzhou 310018, China}\affiliation{Institute of Modern
Physics, Chinese Academy of Sciences, Lanzhou 730000, China}

\author{Lei Zhang}
\affiliation{School of Information Engineering, Hangzhou Dianzi
University, Hangzhou 310018, China}

\author{Gao-Chan Yong}
\affiliation{Institute of Modern Physics, Chinese Academy of
Sciences, Lanzhou 730000, China} \affiliation{University of Chinese
Academy of Sciences, Beijing 100049, China}

\author{Zi-Yu Liu}
\affiliation{College of Physics and  Electronic Engineering,
Xianyang Normal University, Xianyang 712000, China}

\author{Wei Zuo}
\affiliation{Institute of Modern Physics, Chinese Academy of
Sciences, Lanzhou 730000, China} \affiliation{University of Chinese
Academy of Sciences, Beijing 100049, China}

\begin{abstract}
Based on the isospin dependent transport model IBUU, the pion
production and its absorption are thoroughly studied in the central
collision of $Au+Au$ at a beam energy of 400 MeV/nucleon. It is
found that the pions are firstly produced by the hard $\Delta$ decay
at the average density around $1.75\rho_{0}$, whereas about 18\% of
them are absorbed absolutely in the subsequent inelastic collisions.
For the free pions observed, more than half of them have been
scattered for one or more times before they are free from matter.
And the more scattering numbers of pions, the higher the momentum
they possess. These pions, due to longer time of their existence in
high density nuclear matter, carry more information about the
symmetry energy of the nuclear matter at high densities.

\end{abstract}

\maketitle

\section{Introduction}

Heavy-ion collisions (HIC) offer a unique possibility to study bulk
properties of hot and compressed nuclear matter. One of the main
goals of such study is to determine the density-dependence of the
symmetry energy (SE) at high densities \cite{li08,bar05,Colonna20}.
The symmetry energy plays essential roles in understanding a number
of physical phenomena and processes. However it can not be directly
measured in experiments and has to be extracted from observables
which are sensitive to the symmetry energy\cite{Gautam11}.

Pion production is a dominant feature in heavy-ion collisions at
intermediate energies\cite{liba05,lo18,Zhang17,cozma17}. In the
collisions near the pion-production threshold the nuclear matter can
be compressed up to about 2 times normal density $\rho_{0}$ before
it expands again. During the compression stage the pions are
produced mainly from the decay of the $\Delta$ resonances
created\cite{liba15,song15,zhangyx18}. Therefore the pion production
is considered to be important for extracting the information of the
properties of the nuclear matter at high densities\cite{cozma16}.
Since the charged pion ratio in heavy-ion collisions was first
suggested in Ref.~\cite{li02}, pion production has attracted much
attention in pinning down the density dependence of the symmetry
energy. During the last decade, a lot of pionic observables have
been proposed as promising probes, such as the collective pion
flows\cite{liq06,liu18,gao18}, the cone-azimuthal emission of
charged pions\cite{gao13}, etc. However, comparing the theoretical
results of the $\pi^-/\pi^+$ ratio with the experimental data,
analyses came to rather conflicting conclusions on the stiffness of
the nuclear symmetry energy at high
densities\cite{xiao09,xie13,feng10}.

In the reactions near the threshold of pion-production the pions are
mainly the decay products of the $\Delta$ resonances. However most
of the them can be absorbed into $\Delta$ resonances because of the
inelastic $\pi$$N$ collisions, and then may decay into pions
frequently, i.e. scattering process\cite{Bass93,li94,lv17}. The
detail of the scattering process plays an important role for
extracting the information of the symmetry energy at different
densities. At the same time, it is commonly known that pions
observed are produced at high densities in heavy-ion collisions. The
quantitative study of the density at which pion produces, as well as
the pion absorption or its scattering, is seldom reported yet. To
further extract the information of compressed matter by pionic
observables, it is necessary to perform a detailed analysis of the
pions production and its absorption in HIC.

However, the information of the pion absorption or its scattering
process can not be extracted in experiments so far but can be
obtained in theoretical calculations. In this paper, a detailed
statistical investigation of all the inelastic collisions in the
reaction of Au + Au at 400 MeV/nucleon beam energy is performed. All
the inelastic collisions are recorded and investigated
statistically. Further more, the charged pions at the freeze out are
categorized by their production and re-scattering processes. The
analysis shows the pions are indeed produced at high densities.
Moreover about 60\% of the pions observed have re-absorption and
re-decay processes after they were produced first time, and due to
longer time of their existance in high density nuclear matter, they
carry more substantial information of the high-density behavior of
the symmetry energy than those without any scattering process, which
are produced from hard $\Delta$ decay directly and free from matter
immediately.

\section{The theoretical model}

In the past decade the the isospin-dependent Boltzmann-Uehling-
Uhlenbeck transport model (IBUU) have been very successful in
describing the dynamical evolution of nucleons in phase space, as
well as the reaction dynamics of heavy-ion
collisions\cite{li00,zhangl12,xu18,yang19}. The present IBUU
transport model originating from the IBUU04 model can describe the
time evolution of the single-particle phase-space distribution
function,
\begin{eqnarray}
\frac{\partial f}{\partial
t}+\nabla_{\overrightarrow{p}}E-\nabla_{\overrightarrow{R}}f=I_{c},
\label{buu}
\end{eqnarray}
where $I_{c}$ is the collision item and $f(\vec{r},\vec{p},t)$ is
the phase-space distribution function which denotes the probability
of finding a particle at time $t$ with momentum $\vec{p}$ at
position $\vec{r}$. $E$ denotes the total energy, ie, $E=E_{kin}+U$.
$U$ is the mean-field potential of the single particle and can be
expressed as\cite{Das03}
\begin{eqnarray}
U(\rho, \delta, \textbf{p},\tau)
=A_u(x)\frac{\rho_{\tau^\prime}}{\rho_0}+A_l(x)\frac{\rho_{\tau}}{\rho_0}\nonumber\\
+B\left(\frac{\rho}{\rho_0}\right)^\sigma\left(1-x\delta^2\right)\nonumber
-8x\tau\frac{B}{\sigma+1}\frac{\rho^{\sigma-1}}{\rho_0^\sigma}\delta\rho_{\tau^{\prime}}\nonumber\\
+\frac{2C_{\tau,\tau}}{\rho_0}\int{d^3\textbf{p}^{\prime}\frac{f_{\tau}(\textbf{r},
\textbf{p}^{\prime})}{1+\left(\textbf{p}-
\textbf{p}^{\prime}\right)^2/\Lambda^2}}\nonumber\\
+\frac{2C_{\tau,\tau^\prime}}{\rho_0}\int{d^3\textbf{p}^{\prime}\frac{f_{\tau^\prime}(\textbf{r},
\textbf{p}^{\prime})}{1+\left(\textbf{p}-
\textbf{p}^{\prime}\right)^2/\Lambda^2}}, \label{Un}
\end{eqnarray}
where $\tau,\tau^\prime=\pm1/2$ denote the neutron and the proton,
respectively. The variable $x$ denote the stiffness of the symmetry
energy. Varying the $x$, one can get different forms of the symmetry
energy predicted by various many-body theories without changing any
property of symmetric nuclear matter and the value of symmetry
energy at normal density $\rho_0$. The parameters
$A_{u}(x),A_{l}(x)$ are x dependent and defined as
\begin{equation}
A_{u}(x)=-95.98-\frac{2B}{\sigma +1}x,
\end{equation}
\begin{equation}
A_{l}(x)=-120.57+\frac{2B}{%
\sigma +1}x.
\end{equation}
The parameter values are $B=106.35$ MeV, $\sigma$=4/3. $\Lambda
=p_{F}^{0}$ is the nucleon Fermi momentum in
symmetric nuclear matter, $C_{\tau ,\tau ^{\prime }}=-103.4$ MeV and $%
C_{\tau ,\tau }=-11.7$ MeV. The $C_{\tau ,\tau ^{\prime }}$ and
$C_{\tau ,\tau }$ terms are the momentum-dependent interactions of a
nucleon with unlike and like nucleons in the surrounding nuclear
matter. With this potential we can get binding energy -16 MeV and
incompressibility 211 MeV for symmetric nuclear matter and the
symmetry energy 31.5 MeV at saturation density. The resonance
$\Delta$ potential is given by
\begin{eqnarray}
\begin{split}
U^{\Delta^-}&=U_{n},\\
U^{\Delta^0}&=\frac{2}{3}U_{n}+\frac{1}{3}U_{p},\\
U^{\Delta^+}&=\frac{1}{3}U_{n}+\frac{2}{3}U_{p},\\
U^{\Delta^{++}}&=U_{p}.
\end{split}
\end{eqnarray}

In the present work, pions are produced via the decay of $\Delta$
resonance. Near the pion-production threshold, the inelastic
reaction channels as follows are taken into account\cite{Engel94},
\begin{eqnarray}
\begin{split}
NN \rightarrow N\Delta   (hard \Delta production),\\
N\Delta \rightarrow NN   (\Delta absorption),\\
\Delta \rightarrow N\pi  (\Delta decay),\\
N\pi \rightarrow \Delta  (soft \Delta production).\\
\end{split}
\end{eqnarray}
The free inelastic isospin decomposition cross sections are
\begin{equation}
\begin{split}
\sigma^{pp\rightarrow n\Delta^{++}}&=\sigma^{nn\rightarrow p\Delta^{-}}=\sigma_{10}+\frac{1}{2}\sigma_{11},\\
 \sigma^{pp\rightarrow p\Delta^{+}}&=\sigma^{nn\rightarrow n\Delta^{0}}=\frac{3}{2}\sigma_{11},\\
 \sigma^{np\rightarrow p\Delta^{0}}&=\sigma^{np\rightarrow n\Delta^{+}}=\frac{1}{2}\sigma_{11}+\frac{1}{4}\sigma_{10}
\end{split}
\end{equation}
The $\sigma_{II^{'}}$ can be parametrized by
\begin{equation}
    \sigma_{II'}(\sqrt{s})=\frac{\pi(\hbar c )^{2}  }{2p^{2}}\alpha(\frac{p_{r}}{p_{0}})^{\beta}\frac{m_{0}^{2}\Gamma^{2}(q/q_{0})^{3}}
    {(s^{\ast}-m_{0}^{2})^{2}+m_{0}^{2}\Gamma^{2}}.
\end{equation}
Here the $I$ and $I'$ are the initial state and final state isospins
of two nucleons, for explicit details and parameters, see
ref.\cite{VerWest82}. The cross section for the two-body free
inverse reaction can be described by the modified detailed balance,
\begin{equation}
    \sigma_{N\Delta\rightarrow NN}=\frac{m_{\Delta}p_{f}^{2}\sigma_{NN\rightarrow N\Delta}}{2(1+\delta)p_{i}}
    \bigg/\int_{m_{\pi}+m_{N}}^{\sqrt{s}-m_{N}}\frac{dm_{\Delta}}{2\pi}P(m_{\Delta}),
\end{equation}
where $p_{f}$ and $p_{i}$ are the nucleon center of mass momenta in
the $NN$ and $N\Delta$ channels, respectively. $P(m_{\Delta})$ is
the mass function of the $\Delta$ produced in $NN$ collision and can
be defined according to a modified Breit-Wigner
function\cite{Danielewicz91},
\begin{equation}
    P(m_{\Delta})=\frac{p_{f}m_{\Delta}\times4m_{\Delta0}^{2}\Gamma_{\Delta}}
    {(m_{\Delta}^{2}-m_{\Delta0}^{2})^{2}+m_{\Delta0}^{2}\Gamma_{\Delta}^{2}}.
\end{equation}
Here $m_{\Delta0}$ denotes the centroid of the resonance and
$\Gamma_{\Delta}$ is the width of the resonance $\Delta$. Assuming
the $\Delta$ be produced isotropically in the nucleon-nucleon center
of mass, and the decay of $\Delta\rightarrow \pi N$ with an
isotropic angular distribution in the $\Delta$ rest frame, the width
of $\Delta$ resonance can be given in a simplistic
fashion\cite{Cugnon81},
\begin{equation}
    \Gamma_\Delta=\frac{0.47q^{3}}{m_{\pi}^{2}[1+0.6(q/m_{\pi})^{2}]}.
\end{equation}
The $q$ is the pion momentum in the $\Delta$ rest frame and defined
as
\begin{equation}
   q= \sqrt{(\frac{m_{\Delta}^{2}-m_{n}^{2}+m_\pi^{2}}{2m_{\Delta}})^{2} -m_\pi^{2}
   },
\end{equation}
The decay of the resonance into the nucleon and the pion is carried
out by the Monte Carlo method for each time step $dt$ in our
calculation, with the probability as
\begin{equation}
    P_{deacy}=1-exp(-dt\Gamma_{\Delta}/\hbar).
\end{equation}
The meson-baryon interactions in our calculations are treated via
the formation of baryon resonances, and the Breit-Wigner form of
resonance formation can be modified as\cite{li01}
\begin{equation}
    \sigma_{\pi+N}=\sigma_{max}(\frac{q_{0}}{q})^{2}\frac{\frac{1}{4}\Gamma_{\Delta}^{2}}{(m_{\Delta}-m_{\Delta
    0})^{2}+\frac{1}{4}\Gamma_{\Delta}^{2}},
\end{equation}
where $q_{0}$ represents the pion momentum at the centroid
$m_{\Delta 0}$=1.232 GeV of the resonance mass distribution. The
maximum cross sections are given by\cite{Cugnon81,Bertsch88,yong17}
\begin{eqnarray}
\begin{split}
\sigma_{max}^{\pi^{+}p\rightarrow \Delta^{++}}&= \sigma_{max}^{\pi^{-}n\rightarrow \Delta^{-}}=200 ~mb,\\
\sigma_{max}^{\pi^{-}p\rightarrow \Delta^{0}}&= \sigma_{max}^{\pi^{+}n\rightarrow \Delta^{+}}=66.67~mb,\\
\sigma_{max}^{\pi^{0}p\rightarrow \Delta^{+}}&=
\sigma_{max}^{\pi^{0}n\rightarrow \Delta^{0}}=133.33~mb.
\end{split}
\end{eqnarray}

\section{Results and discussions}

Transport theories have been very successful in describing the
reaction dynamics of heavy-ion
collisions\cite{Aichelin93,li05a,yong06,wang18,wang20}. Due to their
strong interaction with the nuclear environment pionic observables
at the freeze out are the result of complex creation and
rescattering processes. In order to obtain detail information of the
pion production and its absorption, we study all the inelastic
collisions in the central collision of $Au+Au$ at a beam energy of
400 MeV/nucleon within the frame work of IBUU.
\begin{figure}[th]
\begin{center}
\includegraphics[width=0.5\textwidth]{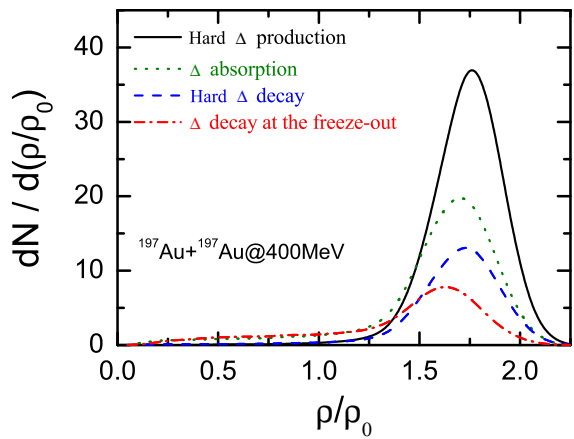}
\end{center}
\caption{Local density distributions of the number of the inelastic
reaction in the central collision of $Au+Au$ at a beam energy of 400
MeV/nucleon.} \label{density}
\end{figure}
Firstly we investigate the numbers of different inelastic reactions
and the densities at which they take place,  which is shown in
Fig.~\ref{density}. By comparing the solid line and the dash line,
it can be seen that only 33\% of hard $\Delta$ can decay into pions
and the rest of them are subsequently absorbed into nucleon without
any decay. Furthermore, due to the low production threshold, pions
are reabsorbed and reproduced quite frequently. About 18\% of the
pions from hard $\Delta$ will be absorbed thoroughly, and the rest
are to be free particles ultimately, but probably having one or more
scattering process ($\pi N\rightarrow \Delta \rightarrow \pi N$)
before they are detected.

In Fig.~\ref{density} we can also see that the reaction $NN
\rightarrow N\Delta$ takes place at the average density around
$1.75\rho_{0}$, and in the almost same range of the density, the
hard $\Delta$ decay into pions. Most of the pions from hard $\Delta$
decay will be scattered in the evolution. Due to the scattering
process, the $\Delta$ decays into  free pions at the freeze-out take
place in a wide density range. Nevertheless, the average density is
up to 1.5$\rho_{0}$. Therefore it is reasonable for pion production
to probe the properties of the nuclear matter at high densities.

\begin{figure}[th]
\begin{center}
\includegraphics[width=0.5\textwidth]{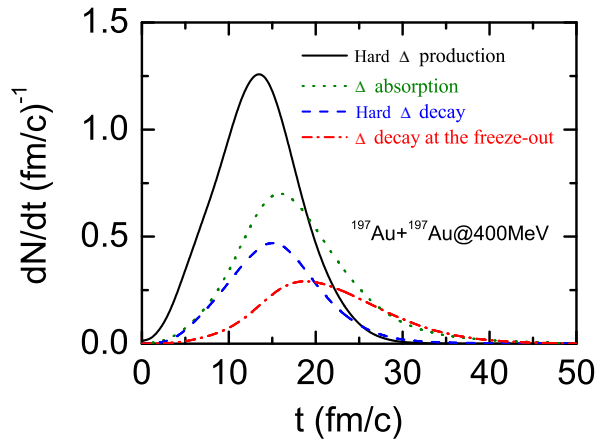}
\end{center}
\caption{Time evolution of the number of the inelastic reaction in
the central collision of $Au+Au$ at a beam energy of 400
MeV/nucleon.} \label{time}
\end{figure}

Fig.~\ref{time} shows the evolution of the inelastic collision
number in the reaction. As common known, the $NN$ inelastic
collisions and the hard $\Delta$ decay  take place in the early
stage. The hard $\Delta$ mostly produces at the time of about
$15fm/c$, and after after 30 $fm/c$, there are almost no new hard
$\Delta$ produced. It can also be estimated Fig.~\ref{time} that the
absorptivity of the pions is about 18\%.

In the following, we focus on the charge pions at the freeze out
because of their advantage in the detector acceptance. With the
analysis of the complex collision history, we classified the free
charge pions,  according to the $\pi N$ rescattering number in their
history.
\begin{figure}[th]
\begin{center}
\includegraphics[width=0.5\textwidth]{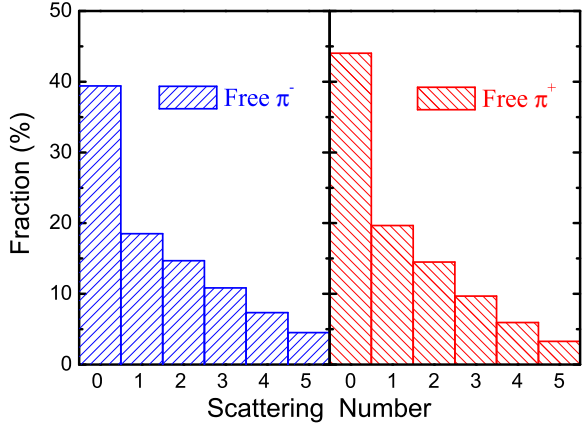}
\end{center}
\caption{Fraction of the different types of free pions categorized
by their scattering numbers.} \label{scattering}
\end{figure}
Fraction of charged free pions originating from specific
rescattering process are plotted in Fig.~\ref{scattering}. Here the
abscissas are the cycle-index of the $\pi N$ scattering of the
collision history for the detectable pions. For example, 0 on the
abscissa is corresponding the pions  without any absorption process
after they are produced from hard $\Delta$ decay, 1 corresponding
the pions have been $\pi N$ scattered for one time, .ie. by the
channel $NN\rightarrow N \Delta\rightarrow NN\pi\rightarrow N
\Delta\rightarrow NN\pi_{free}$, and so on.

It can been seen that most of the free pions have been absorbed into
$\Delta$ and then re-decay into pions after they are firstly
produced by hard $\Delta$ decay. Our calculation shows less than
40\% of the detectable negative pions are seen to freeze out as soon
as they are produced by hard $\Delta$ decay, without any $\pi N$
scattering. It can been also seen that at least 5\% of the negative
pions have been scattered for at least five times before they are
detected.

For positive pions, the fraction of those without $\pi N$ scattering
process is about 44\%, larger than that of the negative ones, as a
result of the Coulomb potential from protons, i.e., negative pions
are attracted to while positive ones are repelled away.

\begin{figure}[th]
\begin{center}
\includegraphics[width=0.5\textwidth]{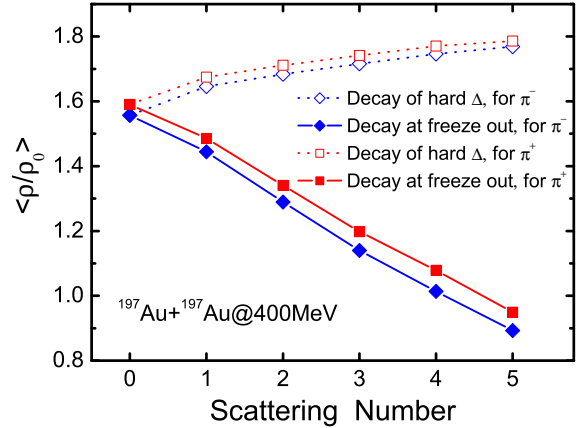}
\end{center}
\caption{The average density where the pion was produced first time
by the hard $\Delta$ decay (first pion formation) and the decay at
the freeze-out (last pion formation), versus the different types of
free pions categorized by their scattering numbers, in the central
collision of $Au+Au$ at 400 MeV/nucleon.} \label{scattering-density}
\end{figure}

We next investigate the pion creation history, and focus on the two
processes. One is the first formation, .ie. the decay of the hard
$\Delta$ into pion, another is last formation, .ie. the $\Delta$
decay into free pion at the freeze-out.
Fig.~\ref{scattering-density} shows the average density at which the
first and last creation take place, for  different categories of
free pion which is categorised in Fig.~\ref{scattering}.  As shown
the average density of the first formation for all categories is
apparently above the normal nuclear density. The pions with more
scattering processes are mostly produced from hard $\Delta$ decay at
higher density. The average density of the last formation of course is
much lower than that of the first formation. Nevertheless, it is still above
the normal nuclear density for most of  categories. The average
densities of the formations of $\pi^{+}$ are little higher than that
of $\pi^{-}$ because of the coulomb potential.

\begin{figure}[th]
\begin{center}
\includegraphics[width=0.5\textwidth]{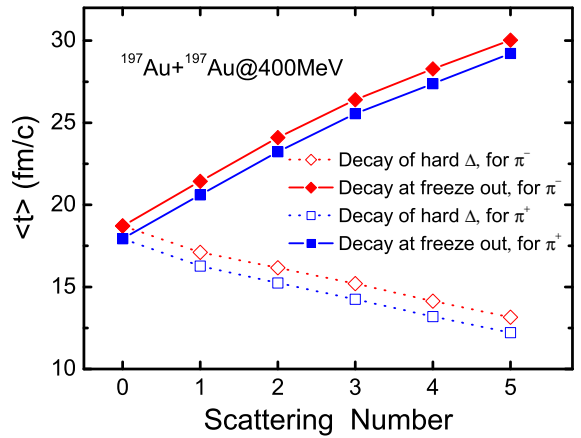}
\end{center}
\caption{The average time when the pion was produced first time by
the hard $\Delta$ decay (first pion formation) and the decay at the
freeze-out (last pion formation), versus the different types of free
pions categorized by their scattering numbers, in the central
collision of $Au+Au$ at 400 MeV/nucleon.} \label{scattering-t}
\end{figure}

Fig.~\ref{scattering-t} shows the average time of the first and the
last formations of the pion, versus the different categories of
pion. One can see that for the categories of the pion with less
scatterings, their first formations take place later. It is not
surprising for the fact that the pions which created from $\Delta$
decay earlier have more probabilities to be scattered, and the whole
scattering process lasts in a larger time range.
\begin{figure}[th]
\begin{center}
\includegraphics[width=0.5\textwidth]{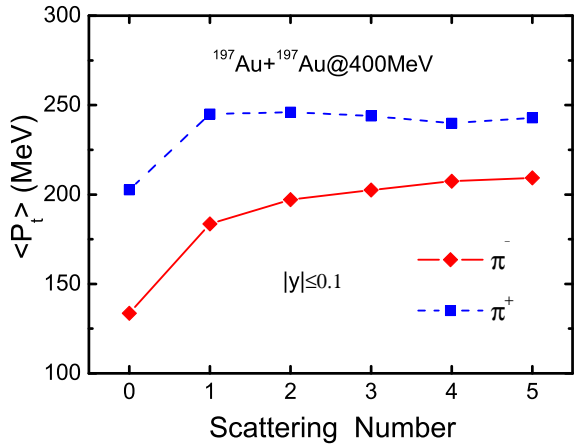}
\end{center}
\caption{The average transverse momentum of the different types of
free pions categorized by their scattering numbers, in the central
collision of $Au+Au$ at a beam energy of 400 MeV/nucleon.}
\label{n-pt}
\end{figure}

\begin{figure}[th]
\begin{center}
\includegraphics[width=0.5\textwidth]{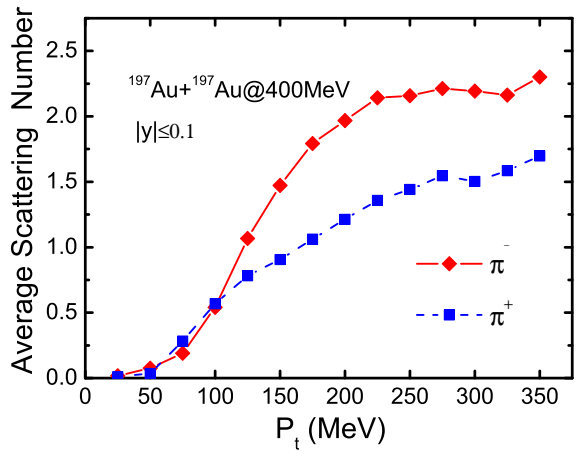}
\end{center}
\caption{The average scattering number in the history of the free
pions as a function of the transverse momentum, in the central
collision of $Au+Au$ at a beam energy of 400 MeV/nucleon.}
\label{pt}
\end{figure}
In Fig.~\ref{n-pt} we investigate the average transverse momentum
$p_{t}$ of the different categories of free pions. One can see that
negative pions with more scattering processes in general have higher
transverse momenta. It also can been seen the average momentum of
the positive pions is obviously higher than that of the negative
pions for each category, due to the coulomb potential.

\begin{figure}[th]
\begin{center}
\includegraphics[width=0.5\textwidth]{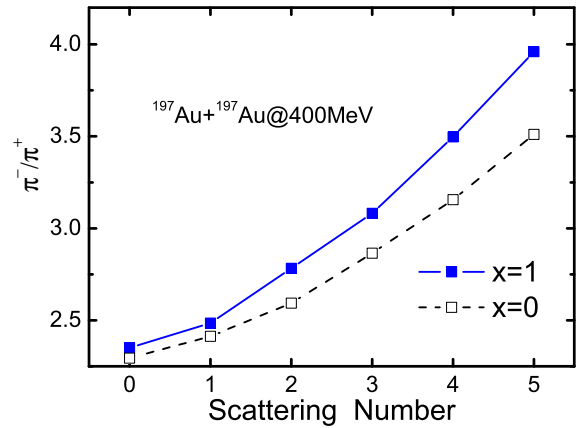}
\end{center}
\caption{The $\pi^{-}/\pi^{+}$ ratio versus the different types of
free pions categorized by their scattering numbers in the central
collision of $Au+Au$ at a beam energy of 400 MeV/nucleon with
different symmetry energies.} \label{ratio}
\end{figure}

We also calculated the average scattering number as shown in
Fig.~\ref{pt} as a function of pion transverse momentum $p_{t}$. The
results show the pions with higher transverse momentum mostly have
more scattering processes, which is consistent with that shown in Fig.~\ref{n-pt}.
It should be noticed that in the transverse momentum spectrum at
small $p_{t}\leqq 50MeV$, the average scattering number is closed to
zero, indicating most of these pions have never been scattered since
they were produced from hard $\Delta$ decay.

Fig.~\ref{ratio} shows the $\pi^{-}/\pi^{+}$ ratio for different
categories of free pions, with soft and stiff symmetry energies,
respectively. We can find that the ratios increase with increasing
scattering number, in both situations. Because of the coulomb
potential, negative pions have more probabilities to be absorbed and
scattered than positive ones. It should also be noticed the effects
of the symmetry energy are apparent for most categories. However for
zero category, i.e. these pions without any scattering process, the
effect appear to be negligible. While with increasing scattering
number, the ratio shows more sensitive to the symmetry energy. In
particular for the pions with five scattering processes, the effect
reach to as much as more than 15\%. The difference of the
sensitivity for different categories is reasonable. The pions
without more scattering processes, are created by hard $\Delta$
decay and freeze out to the detector directly. Whereas for those
pions with more scattering processes, their first formations take
place earlier and last formations take place later. Therefore, the
time of they exit in high density matter is longer, which enhancing
the sensitivity to the stiffness of nuclear symmetry energy at high
densities. This implies that the effect of the symmetry energy is
governed not only by the density where the pions are originated but
also by the time that they spent in high density region during their
whole formation processes.

\section{CONCLUSIONS}

In conclusion, based on the framework of the transport model IBUU,
we investigate the central collision of $Au+Au$ at a beam energy of
400 MeV/nucleon. We analyzed all the inelastic collisions to extract
the information of the pion production and its absorption. The
statistical investigation shows the pions are firstly produced by
the hard $\Delta$ decay at the average density around
$1.75\rho_{0}$. However about 18\% of them are absorbed absolutely
in the subsequent inelastic collisions, thus can not be observed in
the experiment. From categorizing the free pions by their scattering
processes, it is found that most of the free pions have been
scattered for one or more times. The pions with more scattering
numbers have existed in high-density region for longer time. As a
result those pions carry more information of the high-density region
and exhibit significant sensitivity to the symmetry energy.

\section*{Acknowledgments}

The work is supported by the National Natural Science Foundation of
China (11875013, 11775275, 11975282 and 11705041) and the Scientific
Research Program Funded by Shaanxi Provincial Education Department
(Program No.14JK1794).

\end{document}